\documentclass[aps,pra,twocolumn]{revtex4-1}

\usepackage{graphicx}
\usepackage{amsmath}
\usepackage{color}

\begin{document}

\title{Ion crystals in anharmonic traps}

\author{S. R. Bastin}
\affiliation{Department of Physics, Indiana University Purdue University Indianapolis (IUPUI), Indianapolis, Indiana 46202, USA}

\author{Tony E. Lee}
\affiliation{Department of Physics, Indiana University Purdue University Indianapolis (IUPUI), Indianapolis, Indiana 46202, USA}

\date{\today}

\begin{abstract}
There is currently intensive research into creating a large-scale quantum computer with trapped ions. It is well known that for a linear ion crystal in a harmonic potential, the ions near the center are more closely spaced compared to the ions near the ends. This is problematic as the number of ions increases. Here, we consider a linear ion crystal in an anharmonic potential that is purely quartic in position. We find that the ions are more evenly spaced compared to the harmonic case. We develop a variational approach to calculate the properties of the ground state. We also characterize the zigzag transition in an anharmonic potential.
\end{abstract}

\maketitle

\section{Introduction}

A quantum computer can solve certain problems faster than a classical computer \cite{kaye07}. There has been a lot of research into how to construct a large-scale quantum computer. Of the numerous platforms that have been proposed, trapped ions are one of the most promising candidates \cite{cirac95,sorenson99,milburn00}. Trapped-ion quantum computers have demonstrated many quantum algorithms \cite{haeffner08,monz16} and have also demonstrated the ability to store quantum information for long periods of time \cite{harty14}.

\begin{figure}[b]
	\centering
	\includegraphics[width=\columnwidth, trim=0.6in 4.4in 0.9in 4.6in, clip]{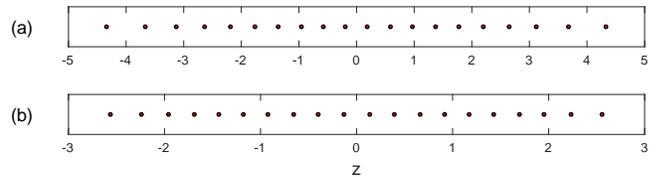}
	\caption{\label{fig:spacing} Ground state crystals of 20 ions in (a) quadratic and (b) quartic potentials. Notice that the ions in the quartic potential are more evenly spaced. In these plots, the position $z$ has been rescaled to be dimensionless.}
\end{figure}

An issue with ion traps is how to scale up to hundreds of qubits. A Paul trap uses a combination of rf and dc voltages to trap the ions \cite{ghosh95}. Typically, the trapping potential is harmonic (quadratic in position) in all three dimensions. The harmonic potential in the axial direction is usually set to be weaker than in the transverse directions so that the ions form a linear string along the axial direction \cite{steane97}. It is well known that the ions near the center of the string are more closely spaced than the ions near the ends [Fig.~\ref{fig:spacing}(a)] \cite{james98}. This bunching in the center is problematic for two reasons. First, it is difficult to focus a laser to individually address the center ions. Second, the bunching in the center causes the string to buckle in a transverse direction \cite{schiffer93,dubin93}; this ``zigzag'' transition occurs because it is energetically favorable for the ions to be displaced in the transverse direction.

This bunching effect is more pronounced as the number of ions increases, which makes it difficult to construct a large-scale quantum computer in a single harmonic trap. Several solutions have been proposed for this problem. One solution is to divide a trap into many separate traps and shuttle ions between the different traps \cite{kielpinski02}. Alternatively, one can connect separate traps via optical cavities \cite{casabone15} or electrostatically \cite{daniilidis09}.

Another solution is to let the axial potential be \mbox{anharmonic} so that the ions are more evenly spaced out. Previous works have shown that a combination of harmonic and anharmonic potentials can evenly space out the ions \cite{lin09,doret12}. In this paper, we use a purely anharmonic axial potential that is quartic in position and analyze the properties of long ion crystals in such a potential. We find that a purely quartic potential leads to much more uniform ion crystals compared to a quadratic potential [Fig.~\ref{fig:spacing}(b)]. We develop a variational approach that allows us to analytically calculate the properties for large system sizes. Then we characterize when the zigzag transition occurs in a quartic potential.

Our paper is outlined as follows. In Sec.~\ref{sec:quadratic}, we review the results for a quadratic (harmonic) potential. In Sec.~\ref{sec:quartic}, we present numerical results for a quartic (anharmonic) potential. In Sec.~\ref{sec:variational}, we present analytical results using a variational approach. In Sec.~\ref{sec:zigzag}, we discuss the zigzag transition.


\section{Review of Quadratic Potential}\label{sec:quadratic}


We briefly review the results for a linear string of ions in a quadratic potential. For now, we ignore the trapping potentials in the transverse directions, since they are assumed to be much stronger than in the axial direction, such that the ions are always at the minimum of the transverse trapping potentials. The potential energy of the system is then \cite{james98}
\begin{equation} \label{potential}
	E=\sum_{i=1}^{N} \frac{1}{2} m \omega^{2} z_{i}^{2} + \sum_{\substack{i,j=1\\i<j}}^{N} \frac{q^{2}}{\left| z_i - z_j \right|},
\end{equation}
where $m$ and $q$ are the mass and charge of an ion, $\omega$ is the axial trap frequency, $z_i$ is the axial position of each ion, and $N$ is the number of ions. For convenience, we rescale the positions $z_i$ using a length scale $\ell=q^2/m\omega^2$. Then the rescaled potential energy is
\begin{equation} \label{potential_scaled}
	E=\sum_{i=1}^{N} \frac{1}{2} z_{i}^{2} + \sum_{\substack{i,j=1\\i<j}}^{N} \frac{1}{\left| z_i - z_j \right|}.
\end{equation}

We are interested in the ground state of Eq.~\eqref{potential_scaled}, i.e., the configuration of $\{z_i\}$ that minimizes the potential energy. When the ions are laser-cooled to sufficiently low temperatures, they form a crystal with this configuration.

Figure~\ref{fig:spacing}(a) shows the ground state for $N=20$. Note that the ions in the center are closer together than those at the ends \cite{james98}. The bunching near the center can be understood in terms of energy considerations. In Eq.~\eqref{potential_scaled}, the first term (trap potential energy) is minimized when all the ions are at the center of trap, whereas the second term (Coulomb energy) is minimized when the ions are spaced far apart. The actual ground state balances these two competing effects. Since the energy from the trapping potential increases with $z$, it is energetically favorable for the ions near the center to be closer together; this reduces the trap potential energy at the expense of an increase of Coulomb energy.


In order to analytically characterize long ion strings, Dubin developed a variational approach \cite{dubin97}. This involves the local-density approximation, i.e., approximating a set of discrete charges by a continuous charge density $n(z)$. The charge density is assumed to be along a line in the axial direction. For a given charge density $n(z)$, the potential energy in Eq.~\eqref{potential_scaled} becomes
\begin{eqnarray} \label{energyN_scaled_simple_quadratic}
	E[n] &= & \int \limits_{-\infty}^{\infty} dz \Bigg\{ \frac{1}{2} z^{2}n(z) + \gamma n(z)^2 \nonumber\\
	&& - \frac {1}{2} n(z) \int \limits_{0}^{\infty} dy \, \ln [y n(z)] \frac{d}{dy} [n(z-y) + n(z+y) ] \Bigg\}, \nonumber\\
\end{eqnarray}
where the first term in the integrand is the energy due to the quadratic trap potential, and the other terms are the Coulomb energy of the chain \cite{dubin97}. $\gamma\approx 0.57721$ is Euler's constant.


Finding the ground state becomes a matter of finding the function $n(z)$ that minimizes Eq.~\eqref{energyN_scaled_simple_quadratic}. Based on the numerical results for the discrete model, Dubin chose the following variational ansatz for $n(z)$:
\begin{equation} \label{variational_quadratic}
	n(z) =
	\begin{cases}
		\frac{3}{4} \frac{N}{L} \left( 1 - \frac{z^2}{L^2} \right) , & |z| < L, \\
		0, & \text{otherwise},
	\end{cases}
\end{equation}
where $L$ is the half-length of the chain and is the free parameter to minimize the energy with respect to \cite{dubin97}. This ansatz (an inverted parabola) says that the density is maximum in the middle, which means that, in the discrete model, the ions are closest together in the middle. Plugging Eq.~\eqref{variational_quadratic} into Eq.~\eqref{energyN_scaled_simple_quadratic} and minimizing with respect to $L$, one obtains \cite{dubin97}
\begin{eqnarray}
L_\text{min}^3&=&3N\left[\gamma-\frac{13}{5}+\ln(6N)\right], \label{eq:Lmin_quadratic}\\
E_\text{min}&=&\frac{3}{10}NL_\text{min}^2. \label{eq:Emin_quadratic}
\end{eqnarray}
We will extend this variational approach to the quartic potential in Sec.~\ref{sec:variational}.

\section{Quartic Potential} \label{sec:quartic}

Now, we consider an ion string in a purely quartic potential. The potential energy in Eq.~\eqref{potential} is modified to:
\begin{equation} \label{potential_quartic}
	E=\sum_{i=1}^{N} \frac{1}{4} a z_{i}^{4} + \sum_{\substack{i,j=1\\i<j}}^{N} \frac{q^{2}}{\left| z_i - z_j \right|},
\end{equation}
where $a$ is the coefficient of the anharmonic quartic term. We assume $a>0$ so that positive ions are trapped. Such a potential can be generated in a segmented ion trap by suitably designing the electrode structure \cite{lin09,doret12}. (In practice, due to manufacturing and voltage tolerances, there will be a small residual quadratic component in the potential. However, as long as the quartic component dominates over the quadratic component for most of the chain, the ground state will be close to the ground state of a purely quartic potential.)

We rescale the positions according to the length scale $\ell=q^2/a$, so the rescaled potential energy becomes:
\begin{equation} \label{potential_quartic_scaled}
	E=\sum_{i=1}^{N} \frac{1}{4} z_{i}^{4} + \sum_{\substack{i,j=1\\i<j}}^{N} \frac{1}{\left| z_i - z_j \right|}.
\end{equation}
We note that other types of anharmonic ion traps have been considered \cite{lin09,akerman10,lee11,home11,doret12,dubin13,cartarius13,maurice15}.

We are again interested in the ground state of Eq.~\eqref{potential_quartic_scaled}. We have found the ground state for up to $N=1000$ using the conjugate gradient method \cite{press07}. Figure~\ref{fig:spacing}(b) shows the ground state for $N=20$. The ions are much more evenly spaced across the chain compared to the quadratic potential in Fig.~\ref{fig:spacing}(a). The edge ions are slightly farther apart than the center ions, but the difference is less than the quadratic case.

\begin{figure}[t]
	\centering
	\includegraphics[width=\columnwidth, trim=0.8in 2.7in 1in 2.9in, clip]{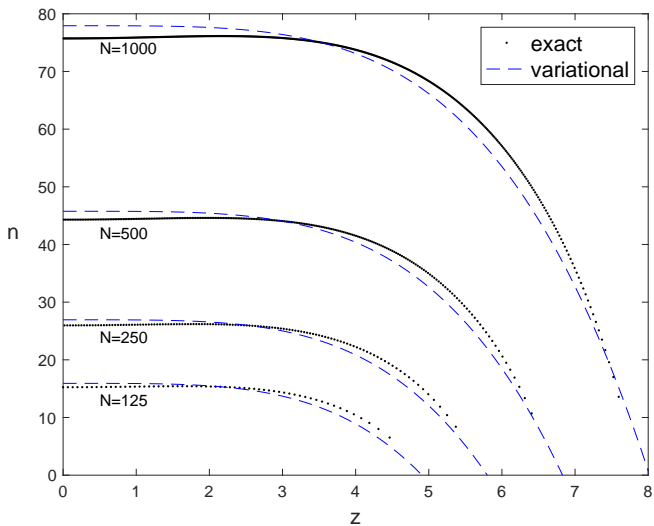}
	\caption{\label{fig:quartic_density} Charge density $n$ as a function of position $z$ for the ground-state ion configuration, comparing the exact results of the discrete model with the variational ansatz for different numbers of ions $N$. Both $n$ and $z$ are in rescaled to be dimensionless.}
\end{figure}

For a configuration of discrete charges, the local charge density can be calculated as
\begin{eqnarray}
n(z_i) = \frac{1}{|z_{i+1}-z_i|},
\end{eqnarray}
since this is the number of ions per unit length \cite{dubin97}. Figure \ref{fig:quartic_density} plots $n(z_i)$ for the ground state of Eq.~\eqref{potential_quartic_scaled} for different $N$. For a large part of the chain, $n(z_i)$ is quite flat. This reflects the fact that most of the ion crystal is evenly spaced. At the ends of the chain, $n(z_i)$ decreases quickly, which means that the ions at the ends are farther apart. Thus, although there is some nonuniformity, it is confined mostly to the ends. The shape of $n(z_i)$ is clearly not an inverted parabola as it was for the quadratic potential [Eq.~\eqref{variational_quadratic}].

The difference between quadratic and quartic results can be understood in terms of energy considerations. A quartic potential is flatter near the center and steeper on the sides compared to a quadratic potential. The ions near the center feel less effect from the trap and therefore space themselves out in order to minimize the Coulomb energy.

An interesting observation from Fig.~\ref{fig:quartic_density} is that the maximum density is not located at the center of the trap ($z=0$), but is located about 1/3 of the distance from the center to the end. This is in contrast to the case of the quadratic potential, where the density is always maximum at the center [Eq.~\eqref{variational_quadratic}].

\begin{figure}[t]
	\centering
	\includegraphics[width=3.2in, trim=0.8in 2.7in 1in 2.9in, clip]{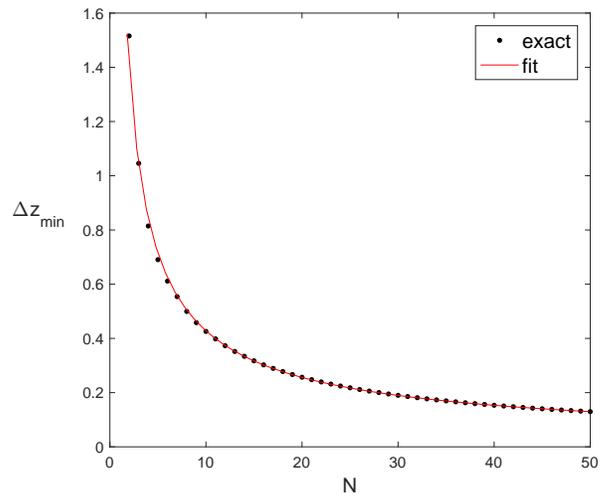}
	\caption{\label{fig:min_separation} Minimum separation between adjacent ions versus number of ions $N$ for a quartic potential, showing the exact results of the discrete model and the fit to a power law. The distance is in rescaled dimensionless units.}
\end{figure}

As $N$ increases, the overall length and density increase. For the quartic potential, the minimum spacing between adjacent ions decreases with $N$ as
\begin{equation}
	\Delta z_\text{min} \approx \frac{2.36}{N^{0.74}}. \label{eq:dz_min}
\end{equation}
The data and the fit are shown in Fig.~\ref{fig:min_separation}. The scaling is different from the quadratic potential, which has $\Delta z_\text{min} \approx 2.02/N^{0.56}$ \cite{james98}. Interestingly, $\Delta z_\text{min}$ decreases faster with $N$ for the quartic potential than the quadratic potential. The reason is that the quartic potential is steeper, so the length of the ion crystal increases slower with $N$.


\section{Variational Approach}\label{sec:variational}

Here, we extend the variational approach of Dubin \cite{dubin97} to the quartic potential. The motivation is that it is very time consuming to numerically find the ground state for large $N$, e.g., more than four days for $N=1000$. The advantage of the variational approach is that it allows us to analytically calculate the properties of the ion crystal.

We make the local-density approximation, i.e., we approximate the discrete charge density with a continuous function $n(z)$. The potential energy is the same as Eq.~\eqref{energyN_scaled_simple_quadratic}, but the first term is modified:
\begin{eqnarray} \label{energyN_scaled_simple_quartic}
	E[n] &= & \int \limits_{-\infty}^{\infty} dz \Bigg\{ \frac{1}{4} z^{4}n(z) + \gamma n(z)^2 \nonumber\\
	&& - \frac {1}{2} n(z) \int \limits_{0}^{\infty} dy \, \ln [y n(z)] \frac{d}{dy} [n(z-y) + n(z+y) ] \Bigg\}, \nonumber\\
\end{eqnarray}


We seek the function $n(z)$ that minimizes Eq.~\eqref{energyN_scaled_simple_quartic}. We choose a variational ansatz motivated by the numerical results of the discrete model (Fig.~\ref{fig:quartic_density}). We find that the charge density for large $N$ is well described by an inverted quartic function:
\begin{equation} \label{variational_quartic}
	n(z) =
	\begin{cases}
		\frac{5}{8} \frac{N}{L} \left( 1 - \frac{z^4}{L^4} \right) , & |z| < L ,\\
		0, & \text{otherwise}.
	\end{cases}
\end{equation}
The normalization in Eq.~\eqref{variational_quartic} is determined by the relation $\int_{-L}^{L}n(z)\,dz=N$, i.e., the total charge should be $N$. There is one free parameter, $L$, which is the half-length of the chain. Note that this ansatz does not capture the fact that the maximum density is not at $z=0$. If desired, one could use a more complicated ansatz with more free parameters to better approximate the ground state.



Plugging our variational ansatz [Eq.~\eqref{variational_quartic}] into the expression for potential energy [Eq.~\eqref{energyN_scaled_simple_quartic}], we obtain:
\begin{equation} \label{energyL}
	E = \frac{1}{36} N L^{4} + \frac{5}{9} \frac{N^{2}}{L} \left[ \gamma + \frac{\pi}{2} - \frac{85}{18} + \ln(10N) \right].
\end{equation}
Then we minimize Eq.~\eqref{energyL} with respect to the free parameter $L$ to obtain:
\begin{eqnarray} \label{halfLength}
	L_\text{min}^{5} &=& 5 N\left[ \gamma + \frac{\pi}{2} - \frac{85}{18} + \ln(10N) \right], \\
	E_\text{min} &=& \frac{5}{36} N L_\text{min}^{4}. \label{energy_N}
\end{eqnarray}
Equation \eqref{energy_N} is the variational estimate of the ground-state energy. Note that these results are different from the quadratic case [Eqs.~\eqref{eq:Lmin_quadratic} and \eqref{eq:Emin_quadratic}].


Figure \ref{fig:quartic_density} shows that the variational ansatz has good agreement with the discrete model in terms of the charge density $n$. This indicates that our calculation of $L_\text{min}$ is correct.

\begin{figure}[t]
	\centering
	\includegraphics[width=3.4in, trim=0.in 3.4in 0.2in 3.4in, clip]{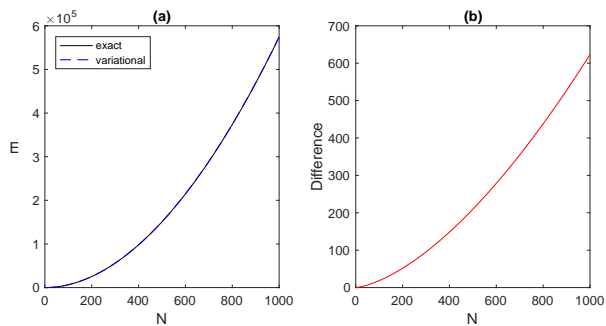}
	\caption{\label{fig:energy} (a) Ground-state energy versus number of ions $N$, comparing the exact results of the discrete model with the variational estimate. (b) Difference between the two curves: $E_\text{exact}-E_\text{variational}$. The energy is in rescaled dimensionless units.}
\end{figure}

Figure \ref{fig:energy} compares the ground-state energy of the discrete model with the variational estimate [Eq.~\eqref{energy_N}]. There is excellent agreement between the two. Note that the variational energy is slightly higher than the true ground-state energy, as expected. (The deviation could be further decreased by using a more complicated variational ansatz with more free parameters.)

We can use the variational results to estimate how $\Delta z_\text{min}$ scales with $N$. From Eq.~\eqref{variational_quartic}, $\Delta z_\text{min}\propto L/N$. From Eq.~\eqref{halfLength}, $L\propto N^\frac{1}{5}$ (ignoring the logarithm that grows slowly with $N$). Thus, the variational approach predicts
\begin{eqnarray}
\Delta z_\text{min}\propto \frac{1}{N^\frac{4}{5}},
\end{eqnarray}
which is close to the numerical result in Eq.~\eqref{eq:dz_min}.


\section{Zigzag transition}\label{sec:zigzag}

Here, we discuss the zigzag transition for a quartic axial potential. To do this, we have to account for possible displacement in the transverse $(x,y)$ directions. We assume that the trap potentials in the transverse directions are still quadratic, but the $x$-potential is stronger than the $y$-potential, so we can ignore the $y$ direction in this discussion \footnote{To be precise, Laplace's equation ($\nabla^2\Phi=0$) implies that a potential cannot be purely quadratic in the $x,y$ directions while being purely quartic in the $z$ direction. However, a potential can be approximately quadratic in the $x,y$ directions while being purely quartic in the $z$ direction. For example: $\Phi=c_1 (x^2-y^2) + c_2 x^4 + c_3 y^4 - 6c_2 x^2z^2 - 6c_3 y^2z^2 + (c_2+c_3) z^4$. If we assume $c_1\gg c_2,c_3$, then for the purposes of calculating the transverse modes, the $x,y$ potentials can be assumed to be quadratic.}. We rewrite the potential energy [Eq.~\eqref{potential_scaled}] to include the $x$ position:
\begin{equation} \label{potential_scaled_transverse}
	E=\sum_{i=1}^{N} \left(\frac{\beta^2}{2} x_{i}^{2} + \frac{1}{4} z_{i}^{4} \right)+ \sum_{\substack{i,j=1\\i<j}}^{N} \frac{1}{\left[(x_i-x_j)^2 + (z_i - z_j)^2 \right]^{1/2}},
\end{equation}
where $\beta$ is the strength of the $x$-potential relative to the $z$-potential.

When $\beta$ is very large, it is energetically favorable for all the ions to have $x_i=0$, i.e., the ground state is a line in the axial direction as we assumed in Sec.~\ref{sec:quartic}. As $\beta$ decreases, at some point, it is energetically favorable for the ions to be displaced in the $x$ direction, and the ground state develops a zigzag shape \cite{schiffer93,dubin93}. We are interested in the critical value, $\beta_c$, at which this zigzag transition occurs. 

The simplest way of calculating $\beta_c$ is to calculate the transverse normal mode frequencies for a linear chain \cite{morigi04}. When $\beta>\beta_c$, all the frequencies are real. When $\beta<\beta_c$, at least one frequency is imaginary; this signals that the ground state is no longer linear.

To calculate the transverse normal modes, we first calculate the $N\times N$ matrix $A$, which is the Hessian of the potential energy, evaluated for the linear ground state \mbox{(with $x_i=0$)} \cite{zhu06}. The elements of $A$ are:
\begin{eqnarray}
A_{mn}&=& \left.\frac{\partial^2 E}{\partial x_m \partial x_n}\right\vert_{\text{lin.~gnd.~state}}, \\
&=&
\begin{cases}
\beta^2 - \sum_{p\neq n} \frac{1}{|z_p-z_n|^3}, & \quad m=n, \\
\frac{1}{|z_m-z_n|^3}, & \quad m\neq n,
\end{cases}
\end{eqnarray}
where the $z_n$ are evaluated using the ground state of Eq.~\eqref{potential_quartic_scaled}. Denoting the eigenvalues of $A$ by $\lambda_n$, the transverse normal mode frequencies are $\sqrt{\lambda_n}$ (note that this is in rescaled units). Thus, $\beta_c$ is found by checking when the smallest eigenvalue reaches zero as $\beta$ decreases. The eigenvector corresponding to this eigenvalue is the ``zigzag'' mode, in which adjacent ions move in opposite transverse directions \cite{zhu06}.

\begin{figure}[t]
	\centering
	\includegraphics[width=3.4in, trim=0.in 3.4in 0.2in 3.4in, clip]{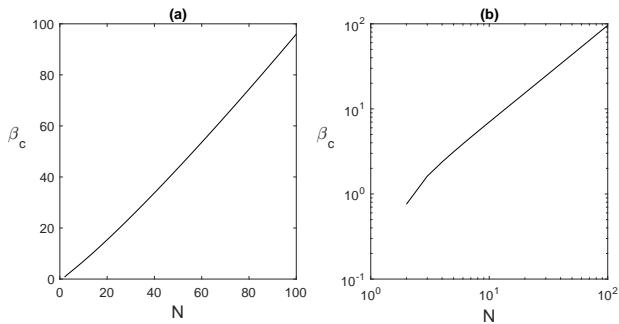}
	\caption{\label{fig:beta_vs_n}Plots of $\beta_c$, the critical strength of the transverse harmonic potential when the zig-zag transition occurs. (a) Linear scale. (b) Log-log scale.}
\end{figure}

Figure \ref{fig:beta_vs_n} shows $\beta_c$ as a function of $N$. We observe the following scaling behavior for large $N$:
\begin{eqnarray}
\beta_c \approx 0.51 N^{1.14}.
\end{eqnarray}
This scaling is different from that for a quadratic axial potential, which has $\beta_c\approx 0.73 N^{0.86}$ \cite{steane97}. It is surprising that $\beta_c$ increases faster with $N$ for the quartic case, despite the ion crystal being more homogeneous. This is due to the fact that $\Delta z_\text{min}$ decreases faster with $N$ for the quartic case than the quadratic case [see Eq.~(10)]. It was shown in Ref.~\cite{totsuji88} that $\beta_c\propto 1/\Delta z^{\frac{3}{2}}$, so $\beta_c$ increases faster with $N$ for the quartic case.

Note that it is difficult to compare the actual values of $\beta_c$ for the two cases because of the difference in units. In the quadratic case, $\beta$ is the ratio between two harmonic frequencies. But in the quartic case, $\beta$ is the ratio between a harmonic frequency and the quartic coefficient. A fairer comparison between the two cases is the following. Suppose we set the strengths of the quadratic or quartic potentials along the $z$ axis such that the ion crystals have the same lengths for the two cases. Then the quartic case will have a larger $\Delta z_\text{min}$ (since the ions are more evenly spaced) and therefore will have a smaller $\beta_c$.

\section{Conclusion}

We have studied the properties of an ion crystal in a purely quartic axial potential. In the future, it would be interesting to develop a variational approach for a potential that contains both quadratic and quartic terms; such potentials have been studied previously in discrete form \cite{lin09}. It would also be interesting to consider higher order potentials like $z^6$ or $z^8$ to see how the scaling behavior of the density or zigzag transition depends on the exponent of the anharmonicity.

Another direction is to investigate whether anharmonicity could help quantum simulation of spin models. The normal modes (either transverse or axial) of a linear ion string can be used to generate effective spin-spin interactions between different ions \cite{porras04}. It is known that when the axial potential is quadratic, the resulting spin-spin interaction decays in distance approximately as a power law. The question then is how a quartic potential would affect the spin-spin interaction. Since ions in a quartic potential are more evenly spaced, the spin-spin interaction may follow a power law more closely.
\vspace{7mm}
\begin{acknowledgments}
S.R.B.~was supported by the Undergraduate Research Opportunities Program (UROP) at IUPUI. The numerical simulations were done on Indiana University's Karst supercomputer.
\end{acknowledgments}

\bibliography{ions}

\end{document}